\documentclass[%
 reprint,
 amsmath,amssymb,
prl
]{revtex4-2}

\usepackage{graphicx}
\usepackage{dcolumn}
\usepackage{bm}
\usepackage{setspace}
\usepackage{amsmath}
\usepackage{amssymb}
\usepackage{amsthm}
\usepackage{physics}
\usepackage{bbold}
\usepackage{bm}
\usepackage{hyperref}
\usepackage{mathrsfs}
\usepackage{verbatim}
\usepackage{mathtools}
\usepackage{amssymb} 
\usepackage{bm}      
\usepackage{mathtools} 
\renewcommand{\selectlanguage}[1]{}
\usepackage[ruled,vlined]{algorithm2e}
\newcommand{\namedsection}[1]{\par\medskip\noindent\textbf{#1:}\ }

\begin{document}

\title{Stochastic Path Compression for Spectral Tensor Networks on Cyclic Graphs}

\author{Ryan T. Grimm, Joel D. Eaves}

\date{\today}

\begin{abstract}
We develop a new approach to compress cyclic tensor networks called stochastic path compression (SPC) that uses an iterative importance sampling procedure to target edges with large bond-dimensions. Closed random walks in SPC form compression pathways that spatially localize large bond-dimensions in the tensor network. Analogous to the phase separation of two immiscible liquids, SPC separates the graph of bond-dimensions into spatially distinct high and low density regions. When combined with our integral decimation algorithm, SPC facilitates the accurate compression of cyclic tensor networks with continuous degrees of freedom. To benchmark and illustrate the methods, we compute the absolute thermodynamics of $q$-state clock models on two-dimensional square lattices and an XY model on a Watts-Strogatz graph, which is a small-world network with random connectivity between spins.
\end{abstract}

\maketitle

  \begin{figure*}
    \centering
    \includegraphics[width=1.0\linewidth]{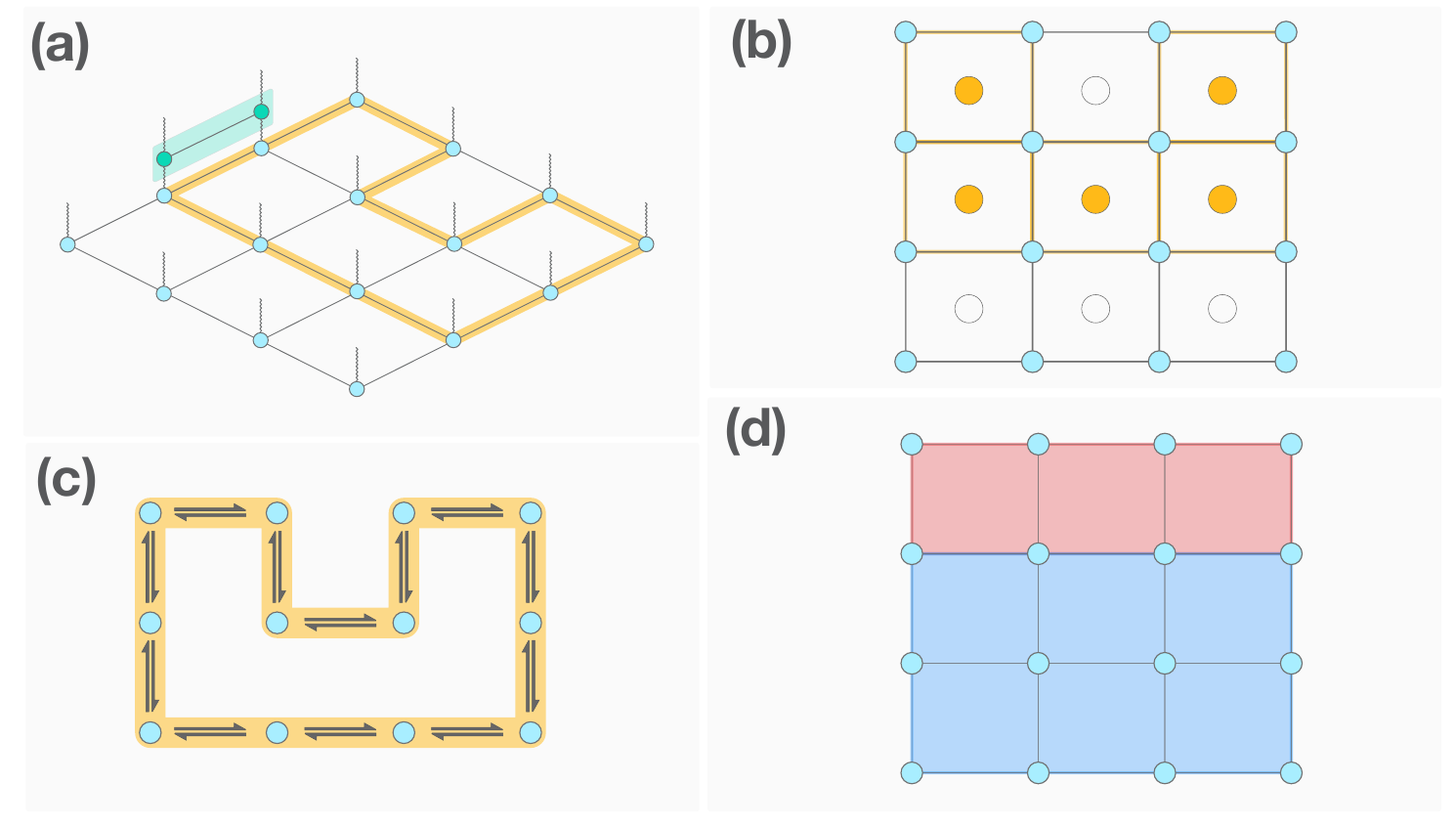}
    \caption{Building a spectral tensor network using integral decimation and compressing it using stochastic path compression. \textbf{(a)} A tensor network on a square lattice and its representation in a compressed form where vertical lines represent the physical bonds that are not contracted over. The Boltzmann weight of a many-body problem on the lattice decomposes into sets of body-ordered gates, and an iterative procedure, starting from a simple product state, forms a computationally tractable partition function. The two-body gate (green rectangle) acts on the network, enlarging the bond-dimension between the two sites. \textbf{(b)} To compress the resulting network, implement a compression cycle. First, draw a random subset of the elementary cycles on the graph to maximize coverage of edges with large bond-dimensions. Generate the compression cycle from symmetric difference (XOR) of the elementary cycles (yellow squares with a filled/unfilled dot denoting the presence of the elementary cycle). \textbf{(c)} In a tensor network with cycles, each edge may need to be compressed more than once. To facilitate this, convert the compression cycle into a directed cycle, where the edge is traversable in both directions. \textbf{(d)} After many iterations, SPC generates an emergent information topology where the network separates into regions of high and low bond-dimensions -- similar to phase separation in two immiscible liquids. The dark line between the red and blue regions represents the interface.     }
    \label{fig:storage_bench}
\end{figure*}
\namedsection{Introduction}
 Tensor network methods (TNMs) exhibit a remarkable capacity to solve a broad class of problems in computational many-body physics and statistical mechanics \cite{orus2014-ky, Chan2011-sb, white1992-mg, strathearn2018-jc, grimm2024-yg, grimm2024-vq}. They compress high-dimensional tensors by decomposing them into sums and products of lower-dimensional tensors, called cores. While the tensors that encode the state of a many-body system are exponentially large and cannot fit in memory, the cores can. They live at the vertices of a graph and are connected to one another through edges. In many cases, the graph's structure mirrors the physical system. The cores have  physical indices that correspond to the system's degrees of freedom (DOF) at a site and virtual dimensions that are contracted over when they share an edge. 
 
TNMs are effective for systems that can be mapped to a one-dimensional topology \cite{white1992-mg, Chan2011-sb} and for systems residing on regular lattices with a small number of states per site \cite{lubasch2014-gk}. The first case corresponds to a set of methods called tree tensor networks (TTNs) \cite{Seitz2023-mc}. Numerous computationally efficient and accurate algorithms exist for constructing, compressing, and contracting TTNs \cite{orus2014-ky, vidal2004-kk}. When systems resist mapping onto a single dimension, the tensor network contains cycles. In these cases projected entangled pair states (PEPS) can perform well in many scenarios \cite{lubasch2014-gk, haghshenas2019-fc, lubasch2014-wx} but they struggle when interactions between cores are long-ranged or when the system is strongly correlated.
 
Devising an accurate, numerically tractable representation of a tensor network is one challenge. Using the network in computations without explosive growth of the bond-dimension---the size of the virtual indices---represents another. Critically, the bond-dimensions must remain small enough to be computationally manageable. Most operations -- multiplication, addition, and gate application -- enlarge the bond-dimension \cite{oseledets2011-ci}. In tensor networks with more than one spatial dimension, a contraction problem emerges where the bond-dimension has a tendency to grow exponentially during contraction. To contain the rapid growth of the bond-dimensions, one applies a compression algorithm that minimizes the squared Euclidean distance (2-norm) between the original network and its approximation as a tensor network with truncated bond-dimensions.  TTNs have a globally optimal compression algorithm \cite{oseledets2011-ci, Seitz2023-mc} that truncates the virtual bond while maintaining a prescribed numerical tolerance \cite{gray2024-gr}. Unfortunately, TTNs do not easily generalize beyond quasi-one dimensional systems without requiring large ($\chi > 100$) bond-dimensions to maintain accuracy \cite{Tagliacozzo2009-oe}. When the network contains cycles, no similar compression procedure is known \cite{tindall2023-th}, which has limited TNMs in their application to more general tensor networks that have cycles, or that have several or continuous DOFs. 

In this letter, we present an algorithm that addresses both limitations and generalizes to graphs with arbitrary topologies. To the extent that the bond-dimension between cores encodes the emergent spatial correlations or entanglement between fluctuations on the lattice, it is natural to employ strategies that sample such fluctuations, like Monte Carlo (MC), to localize the bond-dimension. We develop an iterative sampling procedure called stochastic path compression (SPC). SPC aims to spatially localize ``defective'' or large bond-dimensions in the tensor network by pushing them along random paths in the network, each drawn from a procedure that is inspired by polymer growth in random media. In such an analogy, the cycle pathways are random polymers grown from a MC procedure that biases growth along the edges with the largest bond-dimensions between cores. Once formed, SPC iteratively compresses along this polymer's backbone, forming an ensemble of invasion trajectories that separate large and small bond-dimensions along the polymers grown into the lattice. Repeated application of this procedure leads to a compressed network where large and small bond-dimensions spontaneously localize until they are separated by an interface, which tames the growth of the bond-dimension during tensor contraction. SPC thereby finds tensor compressions iteratively and adaptively in cyclic tensor networks. 


We compute the absolute equilibrium statistical mechanics of lattice systems that are square and that are random, both of which contain cycles. The DOFs $\bm x\equiv \{x_v\}_{v \in \mathbb{V}}$, which may be discrete or continuous, reside on the vertices $\mathbb{V}$ with edges between them $\mathbb{E}$ comprising the  graph $\mathbb{G} \equiv (\mathbb{V}, \mathbb{E})$. The edges $\mathbb{E}$ specify which degrees of freedom interact, and because interactions are local, only connect neighboring vertices. At equilibrium, the Boltzmann factor gives the probability of a configuration $\bm x$ with energy $\mathcal{H}(\bm x)$, $\mathcal{W}(\bm{x}) \equiv  e^{-\beta \mathcal{H}(\bm x)}.$ Specializing discussion to continuous DOFs---discrete DOFs follow similarly---the absolute thermodynamics come from the partition function $\mathcal{Z}  \equiv \int d\bm{x} \; \mathcal{W}(\bm x)$. Differentiation of $\cal Z$ yields absolute entropies, free energies, and specific heats. Because there are exponentially many configurations $\bm x$, the curse of dimensionality typically precludes the direct evaluation of the partition function and absolute thermodynamic quantities. To circumvent this problem, one instead samples configurations from the Boltzmann distribution to obtain relative thermodynamics using, for example, MC methods.

Spectral tensor network methods (STNMs) can enable the evaluation of the partition function and absolute thermodynamic quantities. Spectral tensor networks (STNs) represent multivariate functions as sums and products over tensor-valued univariate functions. Contemporary STNMs generate STNs that are limited to tree-like topologies \cite{bigoni2016-kn, Tang2025-qj}. Tensor networks with cycles, like PEPS, are successful in systems with few discrete DOFs. In this paper, we develop STNMs for graphs with cycles and with many or continuous DOFs. 

To evaluate the partition function through direct integration, we encode the Boltzmann weight as an STN
\begin{equation}
    \mathcal{W}(\bm x) \approx \mathrm{tTr}\left[ \prod_{v \in \mathbb{V}} \mathcal{T}_v(x_v) \right],
    \label{eq:STN_weight}
\end{equation}
where the core $\mathcal{T}_v$ at vertex $v$ is a tensor-valued function,  $\mathrm{tTr}[\cdot]$ is the tensor trace operator that sums over indices on pairs of cores joined by an edge $e$ connecting vertices $(v,v')$ such that $e \equiv (v,v')$, with bond-dimension $\chi_e$ along the edge. Representing the Boltzmann factor as an STN enables storage and integration at computational cost that is linear in the number of DOFs,  $N \equiv |\mathbb{V}|$. The advantage in such a rewriting is that the STN representation in Eq. \ref{eq:STN_weight} is separable. The $N$ dimensional integral over $\cal W$ decomposes into $N$ one-dimensional integrals, turning the interacting problem into a noninteracting one. The result is that the partition function reduces to  $ \mathcal{Z} = \mathrm{tTr}\left[  \prod_{v \in \mathbb{V}} z_v \right]$,
where the single particle partition function is an elementary integral $z_v \equiv \int dx \;\mathcal{T}_v(x)$. 

For simplicity and efficiency, we expand the cores in a set of orthogonal polynomials $\{\phi_k\}$, 
\begin{equation}
    \mathcal{T}_v(x) \equiv \sum_{k = 1}^b \mathcal{C}_{v}^k \phi_k(x),
\end{equation}
where $\mathcal{C}_{v}^k$ are variational tensor-valued coefficients. Hamiltonians contain body-ordered interactions---one-body, two-body, and so on---between DOFs. Given the success of a method we devised called integral decimation (ID) to represent multi-variate functions as one-dimensional STNs \cite{Grimm2025-lx}, we extend that method here to represent multi-variate functions on cyclic graphs.

ID constructs STNs by mapping the Boltzmann factor to a quantum circuit. For a two-body Hamiltonian with the form  \begin{equation}
    \mathcal{H}(\bm x) = \sum_{(v,v')\in \mathbb{E}} H_{v,v'}(x_v, x_{v'}),
\end{equation}
the Boltzmann factor decomposes into a series of low order interactions
\begin{equation}
     \mathcal{W}(\bm{x}) \equiv \prod_{(v,v')\in \mathbb{E}} \mathcal{U}_{v,v'}(x_v,x_{v'}),
\end{equation}
where, drawing language from quantum simulation, the gates are 
\begin{equation}
    \mathcal{U}_{v,v'}(x,x') = e^{-\beta H_{v,v'}(x,x')}.
    \label{eq:gate}
\end{equation}
Applying the gates in sequence to a unit valued tensor network of product states yields the coefficient tensors. Fig. \ref{fig:storage_bench} (a) is an example of applying a two-body gate to the tensor network. The application of a gate enlarges the bond-dimension along the corresponding edge. Compressing after each step controls the expansion of the bond-dimensions in the next section, which makes the underlying ID process possible even for tensor networks with cycles.

\namedsection{Stochastic Path Compression}
For a general weight function the approximation in Eq. \ref{eq:STN_weight} becomes exact as the bond-dimension increases. At finite bond-dimensions, the goal is to minimize $\mathcal{L} = ||\mathcal{W} - \tilde{\mathcal{W}}||_2^2,$ where $\cal W$ is the STN before truncation, $\tilde{\cal W}$ is the network after truncation, and $||\cdot||_2$ denotes the 2-norm. Directly minimizing $\cal L$ across all cores simultaneously leads to a global optimization problem that can be non-convex and subject to the vanishing gradient problem, leading to a computationally prohibitive problem. Minimizing $\cal L$ requires contracting the full network, an example of the contraction problem, which is in a harder computational complexity class for nontrivial cyclic tensor networks than for TTNs. In TTNs, there are methods to find the cores that minimize $\cal L$ without contracting the entire network. When cycles are present in the network, however, there is no tractable algorithm where accuracy is bounded unless one contracts the whole network. Stochastic path compression extends the standard compression algorithm for a matrix product state (MPS) \cite{oseledets2011-ci}. It provides an approximate, but accurate, algorithm that acts on closed, cyclic path $\mathbb{P} \equiv \{v_1, v_2, \dots, v_1\}$, where the paths are embedded into the larger network, or environment, that are the cores not in $\mathbb{P}$. 

A natural method is to partition a graph $\mathbb{G}$ into its cycle basis $\mathcal{C}(\mathbb{G})$, which is related to the set of Eulerian paths---those paths that traverse each edge only once. The cycle basis, in turn, has a compact representation as a configuration of a binary Ising model, where configurations are collections of state variables $c_j \in \{0,1\}$, Fig. \ref{fig:storage_bench} (b). Note that the cycle configuration may contain disconnected loops. The cost of embedding the closed path $\bf{c}$ in the lattice corresponds to the grand canonical Hamiltonian of the microstate
\begin{equation} -\mathcal{E}(\bm c) = \mu |\mathbb{C}(\bm c)| \ + \frac{1}{\chi_{\mathrm{max}}} \sum_{e \in \mathbb{C}(\bm c)} \chi_e   
\end{equation}
where $\mathbb{C}(\bm c)$ denotes the set of edges for the cycle represented by $\bm c$, $e \equiv (v,v')$ is the edge, $\chi_e$ is the bond-dimension along that edge, $\mu$ is the chemical potential that fixes the average path length, and $\chi_{\mathrm{max}}$ is the max bond-dimension. The first term sets the average length of the path, and the second biases each path so it covers edges with high bond-dimensions. SPC adaptively selects paths by drawing $\bm c$ with Metropolis-Hastings \cite{Chib1995-fl} from the fictitious Boltzmann weight  $e^{- \mathcal{E}(\bm c)/\tau}, $ where $\tau$ is a temperature-like meta-parameter that sets the global cost of all paths. 

With a cycle $\mathbb{C}$ chosen, SPC then compresses over an Eulerian circuit on the directed version of the subgraph specified by $\mathbb{C}$, Fig. \ref{fig:storage_bench} (c). We form the directed version of the graph by replacing each undirected edge by a pair of directed edges proceeding in opposite directions. Walking over the directed version is more effective than moving along the cycle as it allows iteration over bonds that do not immediately converge to their optimally compressed form.

Compressing each bond along $v \rightarrow v'$ in sequence yields a locally compressed chain of cores along the path. Each compression operation, which we call a \texttt{push-TSVD} move, has an exact correspondence to a one-dimensional MPS \cite{oseledets2011-ci}. After many iterations, the graph fractionates into regions of high and low bond-dimension density, Fig. \ref{fig:storage_bench} (d).

SPC has similar accuracy to the simple update (SU) method that is common in the PEPS literature. Like SU, SPC ignores correlations in its truncation procedure. Constructing a version of SPC that maintains computational efficiency while considering environmental correlations is challenging and the subject of future work. Belief propagation (BP) \cite{tindall2023-th} provides another lens through which to view SPC. This differs from the typical message passing matrices used for PEPS as the move pushes information along the bond from $v$ to $v'$, where the unitary core at $v$ contains no information. The flow of information along the path allows the spontaneous separation of high and low bond-dimensions. Though, unlike BP, which terminates in the Vidal gauge, SPC does not necessarily approach a unique fixed point. 

\begin{algorithm}[H]
\DontPrintSemicolon
\SetKwInOut{Input}{Input}\SetKwInOut{Output}{Output}

Compute the cycle basis $\mathcal{C}(\mathbb{G})$ of $\mathbb{G}$.\;
Initialize binary vector $\bm{c} \in \{0, 1\}^D$ to a uniform random bit-string, where $D \equiv |\mathcal{C}(\mathbb{G})|$.\;

\For{$i \coloneq 1$ \KwTo $N_{\mathrm{metro}}$}{
    Draw $k \in \{1, \dots, D\}$ uniformly.\;
    $c'_k \coloneq 1 - c_k$\;
    $P_{\mathrm{accept}} \coloneq \min\left(e^{-[\mathcal{E}(\bm{c}') - \mathcal{E}(\bm{c})]/\tau}, 1\right)$\;
    $u \coloneq \mathrm{Uniform}(0, 1)$ \;
    \If{$u < P_{\mathrm{accept}}$}{
        $\bm{c} \leftarrow \bm{c}'$\;
    }
}

Compute an Eulerian path $\mathbb{P}$ on the directed graph formed from $\bm{c}$.\;
\ForEach{$(v, v') \in \mathbb{P}$}{
    Perform a \texttt{push-TSVD} move along $v \rightarrow v'$.\;
}
\caption{Stochastic Path Compression}
\end{algorithm}

\begin{figure}
    \centering
    \includegraphics[width=1.0\linewidth]{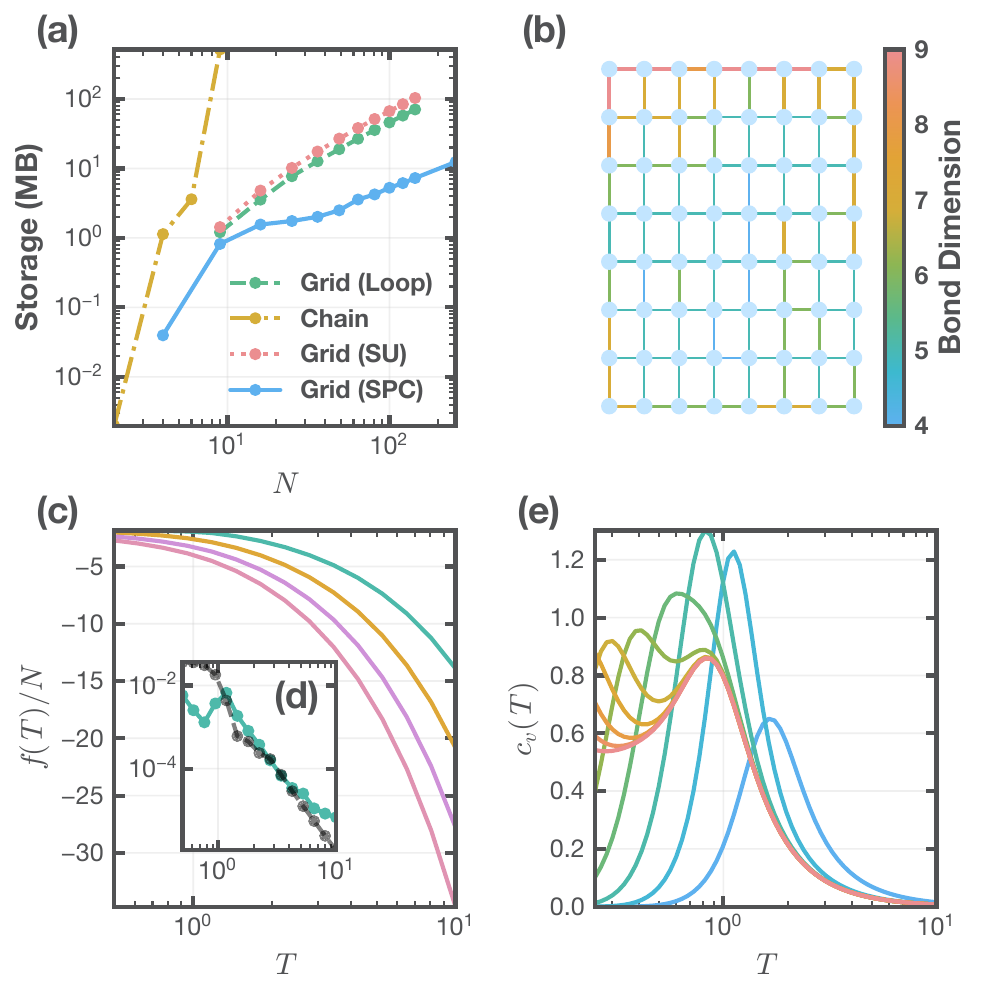}
    \caption{Computing the absolute thermodynamics of the $q$-state clock model on a square lattice using SPC. Energy is in units of $J$, and temperature is in units of $J / k_B.$ We report free energy per site. \textbf{(a)} The storage requirements of a spectral tensor network representation of the Boltzmann distribution for chain and grid representation graphs. We compress the chain with the standard algorithm and the grid with simple update (SU), compression around four-point cycles (Loop), and stochastic path compression (SPC). \textbf{(b)} The final bond-dimension distribution of an $8 \times 8$ lattice. \textbf{(c)} The free energy of a $16 \times 16$ clock model at $q \in \{4, 8, 16, 32\}$ computed with ID and SPC. \textbf{(d)} Relative error in the free energy of an $11 \times 11$ lattice at $q = 4$ compared to the exact transfer matrix solution \cite{levin2007-bs}. The green line shows SPC, and the black line SU. \textbf{(e)} The specific heat of the clock model for $q \in {2, \dots, 11}$ values computed via differentiation of the spectral tensor network.       }
    \label{fig:clock}
\end{figure}

\namedsection{Results \& Discussion}  To benchmark the ability of ID combined with SPC to tackle challenging statistical mechanics problems, we compute the absolute thermodynamics of the two-dimensional clock model. This model allows a systematic increase in the computational complexity because the number of configurational microstates is exponential in $q$. The clock model Hamiltonian on a graph $\mathbb{G} \equiv (\mathbb{V}, \mathbb{E})$ is \begin{equation}
        \mathcal{H}(\bm \theta) = -J\sum_{(v,v') \in \mathbb{E}}  \cos(\theta_v - \theta_{v'}),
        \label{eq:clock}
        \end{equation} where $\bm \theta \equiv \{ \theta_v \}_{v \in \mathbb{V}}$ is the set of angles. The angles take on $q$ discrete values $\theta_v = 2 \pi k/q,$ where $k \in \{1, \dots, q\},$ where $q$ is the number of hours on the clock. Varying the number of hours between $q = 2$ and $q \rightarrow \infty$ interpolates between the binary Ising and continuous XY models. 
        STNs do not scale as the number of DOF, but instead as $\mathcal{O}(Nb^5),$ where $b$ is the number of basis functions per site. Therefore, they can offer a scaling advantage as the number of DOF becomes large. In all examples in this letter, we use Chebyshev polynomials of the first kind for the basis functions.

In all cases, we use our ID algorithm to find the cores and construct the STN. We observe rapid growth in the storage requirements, Fig. \ref{fig:clock} (a), of the one-dimensional STN with the size of the system, limiting the maximum system size to $3\times3$ for reasonable memory constraints of less than one terabyte (TB). To overcome this limitation, we use a grid STN that substantially reduces the growth rate of the storage requirements. Using only the simple update (SU) algorithm from the PEPS literature -- where after applying a gate, we truncate the affected bond to within the SVD cutoff $\epsilon_{\mathrm{SVD}}$ --  reduces the memory requirements by a factor of 100 compared to the one-dimensional STN.

Next, we attempt to further lower the storage requirements by truncating all virtual bonds on the lattice. We compare a naive cycle compression algorithm that truncates bonds around each four-vertex cycle to our SPC method. Cycle compression yields a modest reduction in the storage requirements compared to SU. In contrast, for sufficiently large systems, SPC yields a $90 \%$ reduction in memory requirements. SPC generates an interfacial separation, Fig. \ref{fig:clock} (b), fractionating the bond-dimension distribution between two immiscible fluids. Localizing information in the high-density bond region is substantially more efficient than retaining a uniform bond-dimension. This is only possible in SPC because we allow the bond-dimension to fluctuate. The grid STN with SPC enables us to compute the free energy of up to a $16 \times 16$ grid, Fig. \ref{fig:clock} (c). SU and SPC have similar accuracy with respect to the exact transfer matrix solution to the clock model, Fig. \ref{fig:clock} (d). By including the inverse temperature $\beta$ as auxiliary nodes, all thermodynamic quantities, Fig. \ref{fig:clock} (e), come from first order analytical differentiation (higher orders use finite difference for efficiency) with respect to $\beta$. The free energy requires fewer basis functions ($b = 13$) than the derived thermodynamic quantities do ($b = 61$ for system DOFs and $b = 11$ for temperature). We use $N_{\mathrm{metro}} = 100$ Metropolis steps, a chemical potential of $\mu = -\tau/N$, and a temperature meta parameter of $\tau = 0.01$ for the free energy benchmarks, Fig. \ref{fig:clock} (a-d), and $\tau = 0.1$ for thermodynamics, Fig. \ref{fig:clock} (e).


\begin{figure}
    \centering
    \includegraphics[width=1.0\linewidth]{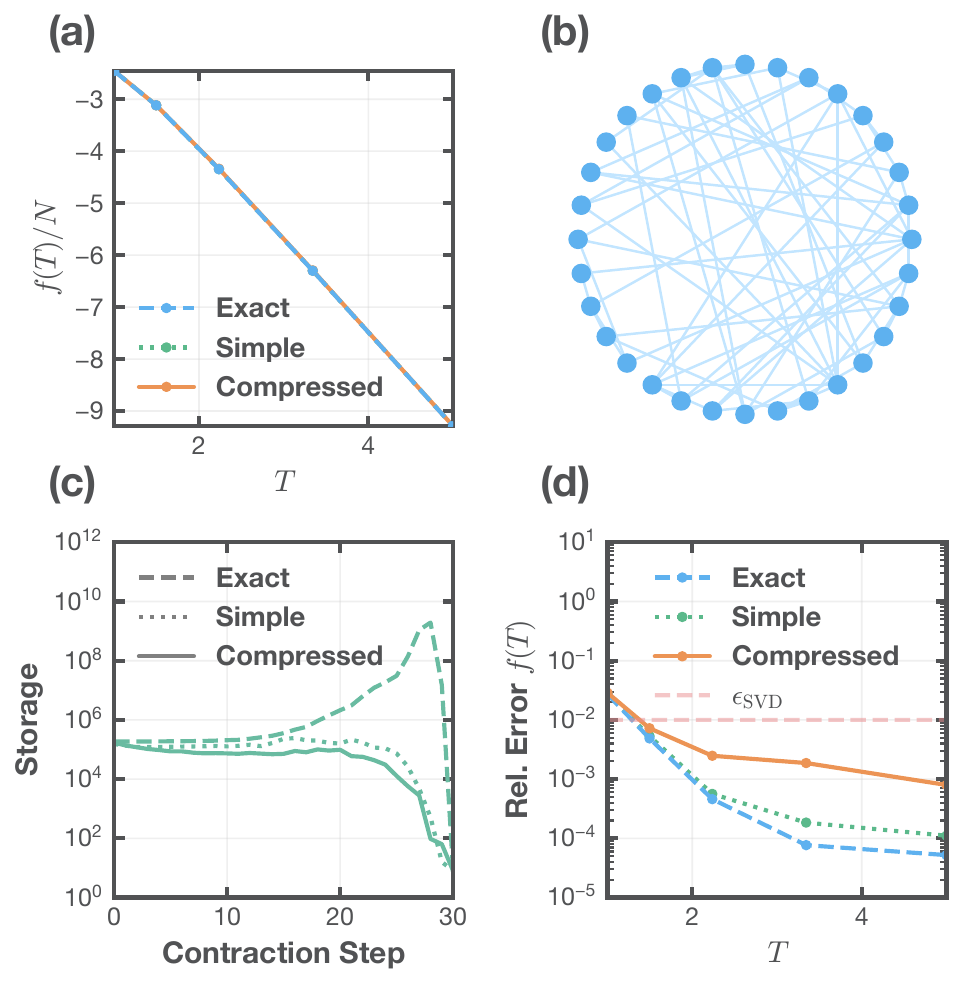}
    \caption{Computing the thermodynamics of the XY model on a Watts-Strogatz graph with $16$, \textbf{(d)}, and $32$, \textbf{(a-c)}, nodes using ID. Energy is in units of $J$, and temperature is in units of $J / k_B.$ We report free energy per site. 
    \textbf{(a)} The free energy computed with three compression modes, where we run SPC after contracting each edge. The modes are exact, no compression applied, simple, SU, and compressed, SPC. \textbf{(b)} The connectivity of the random graph. \textbf{(c)} Storage requirements for each intermediate network during the contraction for $\beta = 1$: without compression, by simple truncation of each edge, and SPC. \textbf{(d)} Relative error compared to a quasi Monte Carlo integration of the partition function. The red dashed line denotes the relative SVD cutoff $\epsilon_{\mathrm{SVD}}$.   }
    \label{fig:watts_strogatz}
\end{figure}

A key feature of our ID and SPC methods is that they are agnostic to the underlying network structure. We illustrate this using a system whose connectivity is not simple, and where the contraction problem becomes much more challenging. To demonstrate these features, we compute the free energy, Fig. \ref{fig:watts_strogatz} (a), of the XY model, Eq. \ref{eq:clock} as $q \rightarrow \infty$, on a random graph, Fig. \ref{fig:watts_strogatz} (b). We use a Watts-Strogatz model \cite{Watts1998-ue}, famous for its ability to model sociophysical phenomena, that generates a random graph with small-world properties by stochastically rewiring connections.

Because the network is not regular, we cannot use the standard boundary contraction algorithm employed in two-dimensional grid networks. Instead, we use the greedy algorithm, developed by Gray and Kourtis \cite{gray2021-pd}. We first construct the STN representation of the Boltzmann weight using the same ID procedure in the previous example. We compare a compressed contraction, where, after each contraction, we compress using SPC ($\tau = 1.0$, $N_{\mathrm{metro}} = 100$, and $\mu = -\tau /N$), to exact contraction where we perform no compression, Fig. \ref{fig:watts_strogatz} (c). The compressed result achieves quantitative agreement, Fig. \ref{fig:watts_strogatz} (d), with respect to Sobol quasi Monte Carlo (QMC) integration using $\sim2^{32}$ samples, exhausting the Sobol grid at $32 \; \mathrm{bit}$ precision. QMC is faster by a square root than MC, so the required number of MC steps, with no importance sampling or other enhancement, would be $2^{64}.$ SPC nearly eliminates growth with respect to the initial storage size of the network, compressing the network by multiple orders of magnitude compared to no compression and by $\sim 80\%$ compared to naive truncation (simple update) of bonds to the cutoff. Except for the lowest temperature $T = 1.0$, the relative error is within the SVD cutoff of $\epsilon_{\mathrm{SVD}} = 10^{-2}$ applied at all stages.

\namedsection{Conclusion}
Algorithms that leverage stochasticity have transformed many computational fields, including, stochastic gradient descent in machine learning \cite{Bottou2010-iq} and randomized linear algebra in numerical computing \cite{Halko2011-xe}. SPC suggests a similar possibility for tensor network methods, where the objective is to generate a compressed approximation to a tensor network that is very good, even if not optimal. Viewing the state of a tensor network as an auxiliary physical system facilitates the development of sampling algorithms developed for those systems.  By letting the bond-dimension fluctuate, we pursue a sampling scheme based on polymer growth in random media that targets regions of large bond-dimensions. After sampling and compressing along many paths, SPC has a tendency to phase separate a tensor network into large and small density domains of bond-dimension. In practical terms, SPC reduces the storage requirements for certain tensor network by up to $90 \%$ and the peak memory required during contraction by up to four orders of magnitude with an accuracy controlled by a cutoff parameter $\epsilon_{SVD}$.  

SPC provides the basis for a new class of algorithms that solve compression and optimization algorithms on embedded, linear tensor networks along adaptively  chosen random paths. Improving the accuracy bounds of SPC is an important direction. Much like the simple update algorithm in the PEPS literature, the present algorithm is only locally optimal and does not consider the effects of the environment of the path when performing the SVD truncations. Incorporating recently developed gauging schemes, such as the belief propagation method by Tindall and Fishman \cite{tindall2023-th}, may enable SPC to operate at similar efficiency while significantly lowering truncation error. Variational energy minimization is another avenue for future exploration. A variant of the density matrix renormalization group (DMRG) that operates on a similar principle to SPC might improve ground state estimations in cyclic tensor networks. 

The problems that are addressable using methods like SPC lie far beyond equilibrium thermodynamics. They include stochastic differential equations, the electronic structure problem, partial differential equations, and quantum relaxation phenomena. SPC and STNs expand the domain of tensor network methods to problems with cyclic topologies and continuous degrees of freedom or large numbers of states. In our work on the Q-ASPEN method applied to quantum relaxation problems, we showed that linear STNs can greatly extend the maximum system size that open quantum systems solvers can access by encoding non-Markovian temporal interactions as an STN \cite{grimm2024-vq}. A future version of Q-ASPEN could use cyclic STNs to treat problems with multiple baths and spatially correlated noise. Further, in much the same way as their discrete counterparts, STNs may provide a more efficient method for circuits in analog quantum computers. Overall, combining SPC with STNs unlocks a new class of stochastic tensor network methods based on tools from statistical mechanics and a range of applications to computationally challenging, continuous systems.

\section*{Acknowledgments}
R.T.G. was supported by the National Science Foundation Graduate Research Fellowship. This material is based upon work supported by the National Science Foundation Graduate Research Fellowship Program under Grant No. (DGE 2040434). Any opinions, findings, and conclusions or recommendations expressed in this material are those of the author(s) and do not necessarily reflect the views of the National Science Foundation. This work was supported by the donors of ACS Petroleum Research Fund under New Directions Grant 68732-ND6. J.D.E. served as Principal Investigator on ACS PRF 68732-ND6 that provided support for R.T.G. This work utilized the Alpine high performance computing resource at the University of Colorado Boulder. Alpine is jointly funded by the University of Colorado Boulder, the University of Colorado Anschutz, and Colorado State University and with support from NSF grants OAC-2201538 and OAC-2322260.

\bibliography{references}
\end{document}